\documentclass[prl,twocolumn,superscriptaddress,notitlepage,floatfix]{revtex4}
\usepackage[sort&compress]{natbib}
\usepackage{graphicx, times}
\usepackage{color}

\begin{document}

\newcommand{\ket}[1]{\ensuremath{\left|{#1}\right\rangle}}
\newcommand{\bra}[1]{\ensuremath{\left\langle{#1}\right|}}

\title{Microwave Photon Detector in Circuit QED}

\date{\today}

\author{G. Romero}
\affiliation{Departamento de
F\'{\i}sica, Universidad de Santiago de Chile, USACH, Casilla 307, Santiago 2, Chile}

\author{J. J. Garc\'{\i}a-Ripoll}
\affiliation{Instituto de F\'{\i}sica Fundamental, CSIC, Serrano 113-bis, 28006 Madrid, Spain}

\author{E. Solano}
\affiliation{Departamento de Qu\'{\i}mica F\'{\i}sica, Universidad del Pa\'{\i}s Vasco - Euskal Herriko Unibertsitatea, Apdo. 644, 48080 Bilbao, Spain}

\begin{abstract}
  In this work we design a metamaterial composed of discrete
  superconducting elements that implements a high-efficiency microwave
  photon detector. Our design consists of a
  microwave guide coupled to an array of metastable quantum circuits,
  whose internal states are irreversibly changed due to the absorption
  of photons. This proposal can be widely applied to different
  physical systems and can be generalized to implement a microwave
  photon counter.
\end{abstract}

\maketitle

Quantum optical photodetection~\cite{mandel95} has occupied a central
role in understanding radiation-matter interactions. It has also
contributed to the development of atomic physics and quantum optics,
with applications to metrology, spectroscopy, and quantum information
processing~\cite{bouwmeester08}. The quantum microwave regime,
originally explored using cavities and
atoms~\cite{walther06,haroche06}, is seeing a novel boost with the
generation of \textit{nonclassical propagating fields}~\cite{houck07}
in circuit quantum electrodynamics
(QED)~\cite{blais04,wallraff04,chiorescu04}. In the last years we have
witnessed a tremendous development of the field of quantum
circuits~\cite{you05,schoelkopf08,clarke08}. These devices are built,
among other things, from superconducting elements, Josephson
junctions, Cooper-pair boxes~\cite{bouchiat98}, SQUID's, microwave
guides and cavities~\cite{blais04,wallraff04,chiorescu04}, all of them
cooled down to the quantum degenerate regime.  Among numerous
applications we may highlight the creation of artificial atoms or
circuits with discrete quantum energy levels, and quantized
charge~\cite{bouchiat98}, flux~\cite{mooij99} or
phase~\cite{martinis85} degrees of freedom. These circuits find
applications not only as quantum bits for quantum information
processing as charge~\cite{vion02,yamamoto03,majer07},
flux~\cite{vanderwal00,chiorescu03} or phase qubits
\cite{berkley03b,simmonds04,steffen06}, but also in the linear and
nolinear manipulation of quantum microwave
fields. In particular we remark the exchange of single photons
between superconducting qubits and
resonators~\cite{cooper04,wallraff04,schuster07,hofheinz08}, the first
theoretical efforts for detecting incoming photons~\cite{helmer07},
the generation of propagating single photons~\cite{houck07} and the
nonlinear effects that arise from the presence of a qubit in a
resonator~\cite{fink08,bishop08}.

While the previous developments represent a successful marriage
between quantum optics and mesoscopic physics, this promising field
suffers from the absence of photodetectors. The existence of such
devices in the optical regime allows a sophisticated analysis and
manipulation of the radiation field which is crucial for quantum
information processing and communication. This includes Bell
inequality experiments, all optical and measurement-based quantum
computing, quantum homodyne tomography, and most important quantum
communication and cryptography~\cite{bouwmeester08}.

\begin{figure}[t]
\centering
\includegraphics[width=0.95\linewidth]{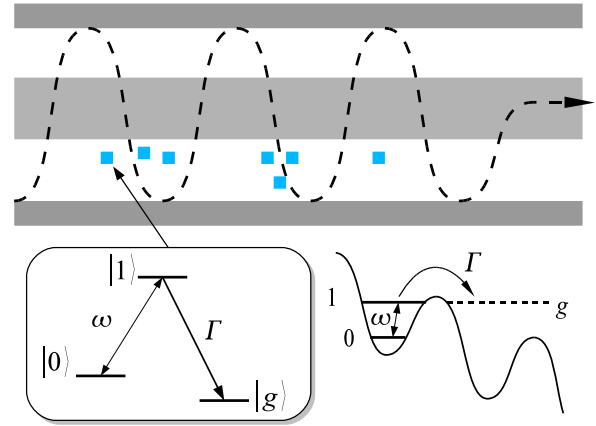}
\caption{Detector design. Sketch showing a one-dimensional waveguide
  (gray) coupled to a set of three-level absorbers (blue squares) in
  arbitrary positions. The microwave field (dashed line) excites
  coherently the state $|0\rangle$ to the upper state $|1\rangle$,
  which decays onto a long-lived stable state. An analogous setup uses
  current-biased Josephson Junctions (CBJJ), in which a washboard
  potential confines two metastable states that can decay into a
  continuum of current states. }
\label{fig:schema}
\end{figure}

There are several challenges to design a microwave photodetector in
circuit QED: i) Available cryogenic linear amplifiers are unable to
resolve the few photon regime. ii) Free-space cross-section between
microwave fields and matter qubits are known to be small. iii) The use
of cavities to enhance the coupling introduces additional problems,
such as the frequency mode matching and the compromise between high-Q
and high reflectivity. iv) The impossibility of performing continuous
measurement without backaction~\cite{helmer07}, which leads to the
problem of synchronizing the detection process with the arrival of the
measured field. We observe that there are related theoretical and
experimental proposals that may solve some but not all of these
problems. Most of them are oriented towards unitary or coherent
evolution of the photon-qubit system, aided by nonlinear dispersive
\cite{fink08,schuster07} or bifurcation effects \cite{siddiqi04}. But
so far, to our knowledge, there has been no implementation of a
microwave photon counter or even a simple microwave photon detector.

\begin{figure}[t]
\centering
\includegraphics[width=0.9\linewidth]{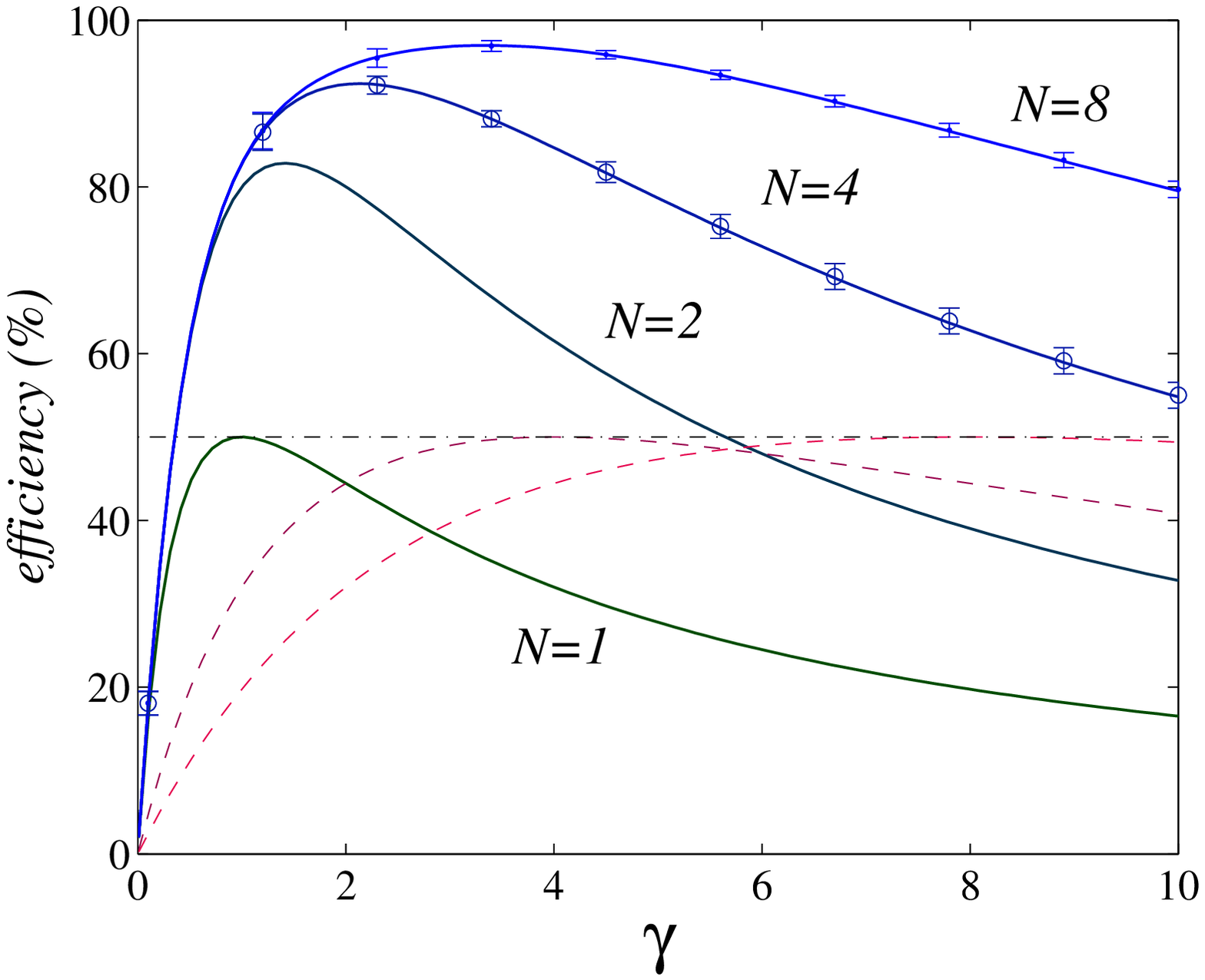}
\includegraphics[width=0.9\linewidth]{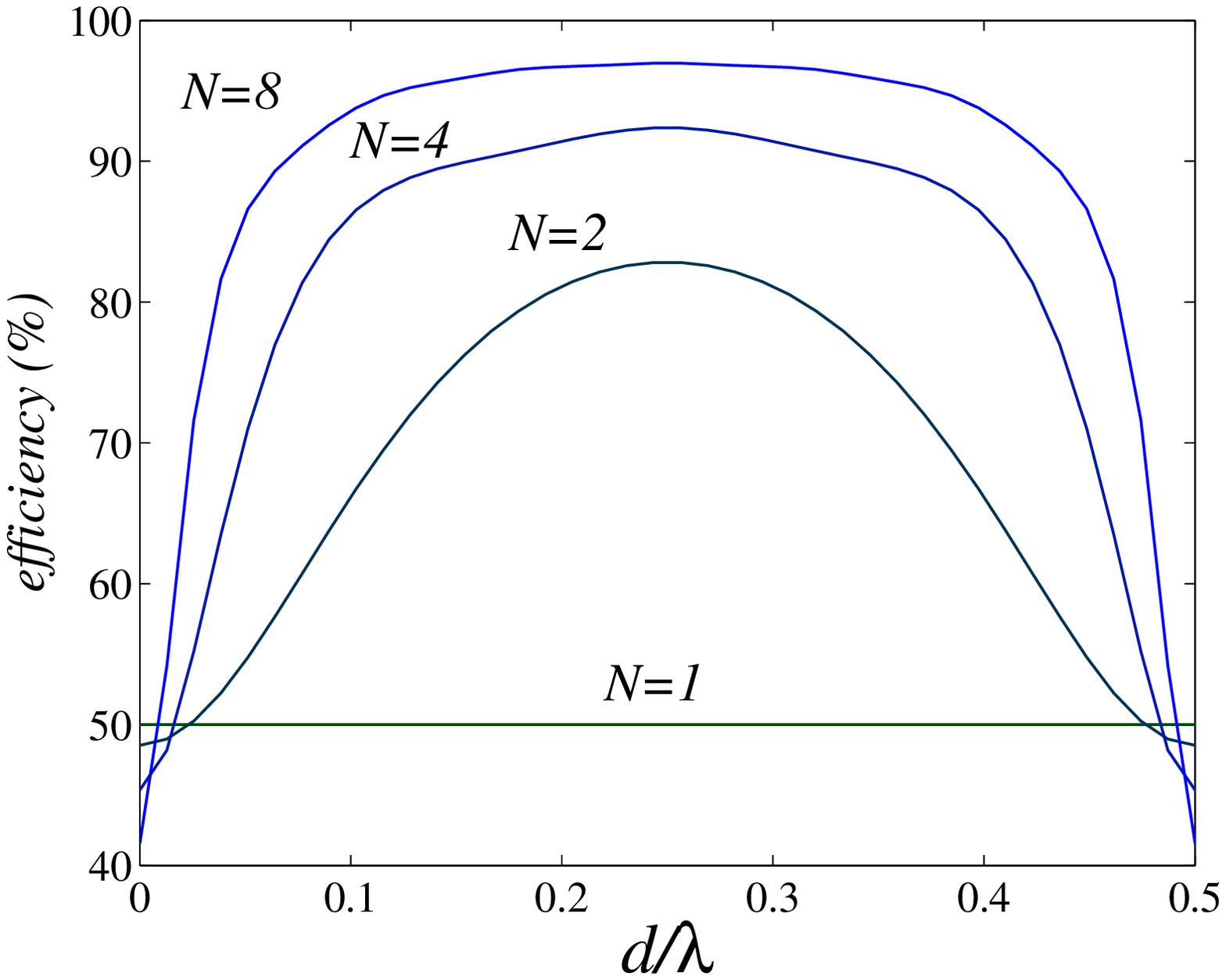}
\caption{(Color online) Detection efficiency when absorbers are on
  resonance (real $\gamma$). (Top) Absorption probability
  vs. effective decay rate $\gamma=\Gamma v_g/V^2$ in dimensionless
  units for a setup with $N=1, 2, 4$ and $8$ qubits (black, green,
  blue, red) either in cluster (dashed) or array (solid) regime.  The
  error bars account for random deviations in the individual absorber
  properties, $\gamma_i,$ of up to 40\%.  When absorbers are close
  together, the efficiency is limited to $50\%,$ while in the other
  case there is no upper limit. (Bottom) Detection efficiency vs.
  separation $d$ in a periodically distributed array of absorbers.}
\label{fig:efficiency}
\end{figure}

Our goal in this work is to design a metamaterial that performs
single-shot microwave photodetection via irreversible absorption of
photons~\cite{patent}. It resembles a photographic film: when a photon
enters the device, there is a large chance that it will be captured,
leaving the system in a stable and mesoscopically distinguishable
state that can be observed \textit{a posteriori}. More precisely, we
propose a general setup based on a one-dimensional waveguide that
passes by a set of photon absorbers [Fig.~\ref{fig:schema}]. We
neither rely on any kind of cavity device nor aim at reaching a
matter-field strong-coupling regime. The absorbers along the waveguide
may be built with bistable quantum circuits similar to the ones used
for implementing qubits. These circuits should be able to capture a
photon and decay from an initial state $\ket{0}$ into a stable state
$\ket{g}.$ In analogy to the photographic film, these irreversible
events constitute the measurement process itself. The final step
consists on counting the number of activated absorbers which is
related to the number of photons in the Fock state which was detected
by the quantum measurement. The counting should be done after the
absorption process has finished, thereby avoiding any disturbing
backaction on the incoming signal. Remark that the detection process
is in this sense passive and does not require any kind of control.

It is possible to develop a simple model for the detection based on
the previous rather general requirements. We model the absorbing
elements as three-level systems, with an internal frequency $\omega$
and a decay rate $\Gamma$ from the excited state $\ket{1}$ to the
final stable state $\ket{g}$, see
Fig.~\ref{fig:schema}. Mathematically, this is described with a master
equation for the density matrix $\rho$ of the absorbers and the
waveguide
\begin{equation}
  \frac{d}{dt}\rho = -\frac{i}{\hbar}[H,\rho] + {\cal L}\rho.
\end{equation}
The Hamiltonian contains terms for the absorbers and the radiation
fields, $\psi_{\mathrm l}$ and $\psi_{\mathrm r}$, propagating left
and right with group velocity $v_g,$ see Methods. The interaction
between both is modeled using a delta-potential of strength $V$
\begin{eqnarray}
  H&=& \sum_i \hbar \omega \ket{1}_i\bra{1} + i\hbar v_g
  \int dx \left[ \psi^\dagger_{\mathrm l}\partial_x\psi_{\mathrm l}
    -\psi^\dagger_{\mathrm r}\partial_x\psi_{\mathrm r}\right]\nonumber\\
  &+& \sum_i \int dx V \delta(x-x_i) \left[ (\psi_{\mathrm{l}} + \psi_{\mathrm{r}})\ket{1}_i\bra{0} +
    \mathrm{H.c.}\right],\label{Heff}
\end{eqnarray}
where $x_i$ and $\ket{0},\ket{1}$ denote the position and the states
of the $i$-th absorber.  The Liouvillian ${\cal L}=\sum_i{\cal L}_i$
has the standard decay terms for each of the absorbing qubits
\begin{equation}
  {\cal L}_i\rho =  \frac{\Gamma}{2}
  \left(2\ket{g}_i\bra{1}\rho\ket{1}_i\bra{g}
    - \ket{1}_i\bra{1}_i\rho - \rho\ket{1}_i\bra{1}\right),  
\end{equation}
and it is proportional to the decay rate $\Gamma.$ The solutions of
this master equation can be found using an equivalent non-Hermitian
Hamiltonian $\bar H = H - i \sum_j\Gamma/2 \ket{1}_j\bra{1},$ that
rules the dynamics of the populations in $\ket{0}$ and $\ket{1}.$ The
norm of the wavefunction is not preserved by this equation, but
precisely the decrease of the norm is the probability that one or more
elements have absorbed a photon.

The simplest scenario that we can consider is a single absorber
coupled to the microwave guide, a problem that has analytical
solutions for any pulse shape. In the limit of long wavepackets it
becomes more convenient to analyze the scattering states of $\bar H.$
For a single absorbing element, these states are characterized by the
intensity of the incoming beam, which we take as unity, and the
intensity of the reflected and trasmitted beams, $|r|^2$ and $|t|^2.$
The associated complex amplitudes are related by the scattering matrix
$T,$ as in $(t, 0)^t = T (1, r)^t.$ The single-absorber transfer
matrix
\begin{equation}
  T = \left(
    \begin{array}{cc}
      1-1/\gamma & -1/\gamma \\ 1/\gamma & 1+1/\gamma
    \end{array}
  \right)
\end{equation}
is a function of a single complex dimensionless parameter
\begin{equation}
  \gamma = (\Gamma - i \delta) v_g/V^2,
\end{equation}
which relates the properties of the circuit: the group velocity in the
waveguide, $v_g,$ the strength of the coupling between the absorbers
and the waveguide, $V,$ and the detuning of the photons from a
characteristic frequency of the absorbers, $\delta=\omega-\omega_\mu.$
The single-photon detection efficiency (absorption probability) is
computed as the amount of radiation which is neither transmitted nor
reflected. In terms of the elements of $T,$ it is given by $\alpha = 1
- (1 + |T_{01}|^2)|T_{11}|^{-2} = 2\gamma(1+\gamma)^{-2}.$

The curve shown in Fig.~2 reveals two regimes. If $\gamma \ll 1$, the
decay channel $\ket{1}\to\ket{g}$ is very slow compared to the time
required for a photon to excite the $\ket{0}\to\ket{1}$ transition,
and only a small fraction of the photons is absorbed. If, on the other
hand, the metastable state $\ket{1}$ decays too fast, $\gamma \gg 1,$
there is a Zeno suppression of the absorption. From
previous formula, the maximal achievable detection efficiency is
50\%, a limit reached by tuning the single absorber on resonance with
the microwave field. We conjecture that this may be a fundamental
limit for any setup involving a single point-like absorber and no
time-dependent external control.

A natural expectation would be that clustering many absorbers inside
the waveguide increases the detection efficiency. As shown in
Fig.~\ref{fig:efficiency}, this is not true. If we have a cluster of
$N$ identical absorbers close together, we can compute the detection
efficiency using the same formula but with the scattering matrix
$T_{\mathrm{cluster}} = T^N.$ As far as the cluster size is smaller
than a wavelength, the setup will be limited to a 50\% maximum
efficiency. There is a simple explanation for this. Since the cluster
size is small, the photon sees the group of absorbers as a single
element with a larger decay rate, $N\Gamma.$ This renormalization just
shifts the location of the optimal working point, leaving the maximum
efficiency untouched.

\begin{figure}
\centering
\includegraphics[width=\linewidth]{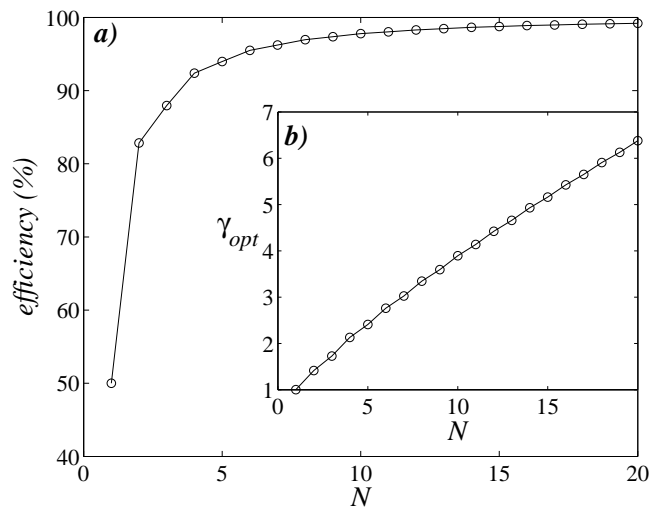}
\caption{(a) Maximal achievable single-photon detection efficiency as
  a function of the number of qubits along the microwave guide. (b)
  Optimal working parameters vs. number of qubits.}
\label{fig:optimal}
\end{figure}

The main result is that we can indeed increase the absorption
efficiency by separating the absorbing elements a fixed distance $d$
longitudinally along the waveguide. The total scattering matrix for
the array is given by
\begin{equation}
  T_{\mathrm{array}} = \prod_{j=1}^N e^{-i 2 \pi \sigma^z d / \lambda} T_j ,
\end{equation}
where $\sigma^z$ is a Pauli matrix and the scattering of each absorber
may be different. In this case the microwave pulse does no longer see
the the detection array as a big particle, and we obtain an collective
enhacement of the absorption probability. Remarkably, an arbitrarily
high detection efficiency can be reached by increasing the number of
absorbers and tuning their separation $d.$ Already with two and three
qubits we can achieve 80\% and 90\% detection efficiency, see
Figs.~\ref{fig:efficiency} and \ref{fig:optimal}. Furthermore, as we
have seen numerically, the more qubits we have, the less sensitive the
whole setup becomes to the experimental parameters, see
Fig.~\ref{fig:efficiency}. This shows that the proposed setup is both
robust and scalable.

The previous analysis is of a general kind. It only requires a
coupling between a waveguide and absorbers that may capture a photon
and irreversibly decay to one or more stable states. A practical
implementation of our setup, which does not require strong coupling or
cavities, consists on a coplanar coaxial waveguide coupled to a number
of current-biased Josephson junctions (CBJJ)
\cite{martinis85,berkley03b}. We will now sketch the microscopic
derivation of Eq.~(\ref{Heff}) for this setup and relate the
efficiency to the parameters of the circuit.

First of all, since the CBJJs are described by a washboard potential
for the phase degree of freedom, we can identify $\ket{0}$ and
$\ket{1}$ with the two lowest metastable levels in a local minimum,
see Fig.~\ref{fig:schema}. The energy levels around such a minimum are
well described by a harmonic oscillator with a frequency $\omega$ that
depends on the bias current. Furthermore, these levels have finite
lifetimes before they decay into the continuum of current states, but
since that the decay rate of state $\ket{0}$, $\Gamma_0,$ can be made
1000 times smaller~\cite{martinis85} than that of $\ket{1}$,
$\Gamma_1,$ we will approximate $\Gamma_0\simeq 0, \Gamma_1\equiv
\Gamma.$

The microwave guide is described by a Lagrangian~\cite{blais04}
\begin{equation}
  L = \int dx \left[ \frac{l}{2}(\partial_tQ)^2 - \frac{1}{2c}(\partial_xQ)^2
    \right],
\end{equation}
where $l$ and $c$ are the inductance and capacitances per unit length.
The quantization of the charge field $Q$ introduces Fock operators
$a_p$ associated to the normal modes of the line. If we assume
periodic boundary conditions, then
$w_p(x,t)=\exp[i(px-\omega_pt)]/L^{-1/2},$ where $L$ is the length of
the waveguide and the dispersion relation is $\omega_p =
v_g|p|=|p|/\sqrt{cl}.$ When the relevant modes of the electromagnetic
field are concentrated in a small interval ${\cal B}$ around a central
momentum $p_0,$ we can introduce right and left moving fields
$\psi_r(x,t) = \sum_{p\in {\cal B}} a_pw_p(x,t),$ and $\psi_l(x,t) = \sum_{p\in{\cal B}} a_{-p}w_{-p}(x,t),$
and approximate the waveguide Hamiltonian as $H = \sum_p \omega_p
a_p^\dagger a_p,$ which corresponds to the first line in
Eq.~(\ref{Heff}).

Finally, for the interaction between the absorbers and the guide we use a
capacitive coupling in the dipole approximation~\cite{blais04}. The
corresponding Hamiltonian has the form
\begin{equation}
  H_{\rm int} = \frac{C_g}{C_g + C_J} \sqrt{\frac{\hbar\omega_\mu}{c}} (
  \psi_r + \psi_l + \mathrm{H.c.}) \frac{e}{\alpha}(a+a^\dagger),
\end{equation}
The first fraction depends on the capacitances of the junction and of
the gate between the junction and the microwave, $C_J$ and $C_g$,
respectively. The second term gives the strength of the electric
potential inside the waveguide and it is proportional to the
fields. Finally, the third term is just the charge operator for the
CBJJ expressed using harmonic approximation around a minimum of the
washboard potential. In particular, $a\simeq \ket{0}\bra{1}$ and
$\alpha^2 = E_C/\hbar\omega$ is the dimensionless parameter of this
oscillator, expressed in terms of the junction capacitance,
$E_C=(2e)^2/C_J$, and the plasma frequency. Note that when we combine
all constants to form the interaction strength $V$ there is not
explicit dependence on the length of the microwave
guide. Qualitatively, while in cavity experiment the qubit only sees a
small fraction of the standing waves with which it interacts, in our
setup each absorber gets to see the whole of the photon wavepacket
after a long enough time.

In terms of the microscopic model, it is possible to write
the parameter that determines the detector efficiency as follows
\begin{equation}
  \gamma = \frac{\alpha^2}{c_{12}} \frac{\hbar}{e Z_0} \frac{\Gamma_1-i(\omega-\omega_\mu)}{\omega_\mu},
 \label{gammaefficiency}
\end{equation}
where we have introduced the dimensionless constant
$c_{12}=C_g/(C_g+C_J).$ It is evident from Eq.~(\ref{gammaefficiency})
that, in order to optimize the efficiency, we have several
experimental knobs to play with. In particular we have considered the
following values, close to current experiments\cite{berkley03b} a
junction capacitance of $C_J=4.8$pF, $c_{12}=0.13$ and $\omega=5$
GHz. Putting the numbers together, and letting the waveguide impedance
oscillate between $10$ and $100~\Omega$, the optimal operation point
for a single junction gives a necessary decay rate
$\Gamma\simeq10-100$ MHz. Increasing $C_g$ by a factor 2 triples the
optimal decay rate, $\Gamma\simeq30-300$ MHz.

Our proposal has the following potential limitations and
imperfections. First, the bandwidth of the detected photons has to be
small compared to the time required to absorb a photon, roughly
proportional to $1/\Gamma.$ Second, the efficiency might be limited by
errors in the discrimination of the state $\ket{g}$ but these effects
are currently negligible~\cite{hofheinz08}. Third, dark counts due to
the decay of the state $\ket{0}$ can be corrected by calibrating
$\Gamma_0$ and postprocessing the measurement statistics. Fourth,
fluctuations in the relative energies of states $\ket{0}$ and
$\ket{1},$ also called dephasing, are mathematically equivalent to an
enlargement of the incoming signal bandwidth by a few megahertz and
should be taken into account in the choice of parameters. Finally and
most important, unknown many-body effects cause the non-radiative
decay process $1\to0,$ which may manifest in the loss of photons while
they are being absorbed. In current experiments\cite{hofheinz08} this
happens with a rate of a few megahertzs, so that it would only affect
very long wavepackets.

A natural extension of our design is the implementation of a photon
counter. This only requires a number of absorbers large enough to
capture all the photons that have to be detected and
counted. Furthermore, while we have chosen three-level systems and
CBJJs for illustrative purposes, the same goal may be achieve using
other level schemes and quantum circuits that can absorb photons and
irreversible decay into long lived and easily detectable states.

The authors thank useful feedback from P. Bertet, P. Delsing,
D. Esteve, M. Hofheinz, J. Martinis, G. Johansson, M. Mariantoni, V. Shumeiko,
D. Vion, F. Wilhelm and C. Wilson. G.R. acknowledges financial support
from CONICYT grants and PBCT-Red 21, and hospitality from Univ. del
Pa\'{\i}s Vasco and Univ. Complutense de Madrid. J.J.G.-R. received
support from Spanish Ramon y Cajal program, and projects FIS2006-04885
and CAM-UCM/910758. E.S. thanks support from Ikerbasque Foundation,
UPV-EHU Grant GIU07/40, and EU project EuroSQIP.

\end{document}